\documentclass[twocolumn,showpacs,preprintnumbers,amsmath,amssymb]{revtex4}
\usepackage{graphicx}
\usepackage{float}
\usepackage[usenames,dvipsnames]{color}
\begin{document}
%\preprint{Draft PRD version 2}
   \title{Where is $\hbar$ Hiding in Entropic Gravity}
   \author{Pisin Chen$^{1,2,3,4}$}
     \email{chen@slac.stanford.edu}
   \author{Chiao-Hsuan Wang$^{1,3}$}
     \email{chwang71@umd.edu}
     \affiliation{%
1. Department of Physics, National Taiwan University, Taipei, Taiwan 10617\\
2. Graduate Institute of Astrophysics, National Taiwan University, Taipei, Taiwan 10617\\
3. Leung Center for Cosmology and Particle Astrophysics, National Taiwan University, Taipei, Taiwan 10617\\
4. Kavli Institute for Particle Astrophysics and Cosmology, SLAC National Accelerator Laboratory, Stanford University, Stanford, CA 94305, U.S.A.
}%

%\date{\today}

\begin{abstract}
The entropic gravity scenario recently proposed by Erik Verlinde reproduced the Newton's law of purely classical gravity yet the key assumptions of this approach all have quantum mechanical origins. So one naturally wonders: where is $\hbar$ hiding in entropic gravity? 
To address this question, we first reformulate the entropic derivation of Newton's gravitation force law to address a self-consistent approach to the problem. Next we argue that as the concept of minimal length has been invoked in the Bekenstein entropic derivation, the generalized uncertainty principle (GUP), which is a direct consequence of the minimal length, should be taken into consideration in the entropic interpretation of gravity. Indeed based on GUP it has been demonstrated that the black hole Bekenstein entropy area law must be modified not only in the strong but also in the weak gravity regime where in the weak gravity limit the GUP modified entropy exhibits a logarithmic correction. 
In the weak gravity limit, such a GUP modified entropy exhibits a logarithmic correction term. When applying it to the entropic interpretation, we demonstrate that the resulting gravity force law does include sub-leading order correction terms that depend on $\hbar$. Such deviation from the classical Newton's law may serve as a probe to the validity of the entropic gravity postulate.
\end{abstract}

%PACS numbers
\maketitle

\section{Introduction}

The issue of how gravity and thermodynamics are correlated has been studied for decades, triggered by the seminal discovery by Bekenstein\cite{Bekenstein,Bekenstein2} on the area-law of black hole (BH) entropy and temperature.
After Hawking's discovery of the BH evaporation and the interpretation of its temperature as the thermal temperature of blackbody radiation\cite{Hawking}, considerable efforts have been made to find the statistical interpretation of the proportionality of black hole entropy and its horizon area. See \cite{Srednicki} and \cite{Area Review}, for example, for a review.
By now a well-accepted view is that the black hole entropy is associated with the external thermal state perceived by an observer outside the event horizon who has no access to the BH interior. Namely, the correlation between the degrees of freedom on opposite sides of the horizon results in a mixed state for observation from the outside, i.e., the `entanglement entropy'\cite{EE&BB,BHJacob}, which depends upon the boundary properties and will be discussed more in the later sections of this paper.

The inversion of the logic that describes gravity as an emergent phenomenon was first proposed by Sakharov \cite{Sakharov}, who suggested that gravity is induced by quantum field fluctuations. Invoking the area scaling property of entanglement entropy, Jacobson in 1995 \cite{Jacobson} used basic laws of thermodynamics to derive Einstein equations. In his perspective, Einstein equations are now an equation of state rather than a fundamental theory. More ideas on emergent gravity have been recently proposed (See, for example, \cite{Volovik,Sidoni et al,Padmanabhan,Verlinde}).

Similar to Jacobson's derivation of Einstein equations through thermodynamic, Verlinde treated gravity as an entropic force analogous to the restoring force of a stretched elastic polymer driven by the system's tendency towards the maximization of entropy \cite{Verlinde}, and interestingly the Newton's law of gravitation was shown to arise. To arrive at the Newton's force law of gravity through the first law of thermodynamic $Fdx=TdS$, Verlinde first invoked the Compton wavelength of the test particle to find the change of entropy with respect to its displacement. He then invoked the holographic principle \cite{'t Hooft, Susskind, Bousso} and the equipartition theorem to define the temperature experienced by the test particle. One cannot but notices that all these building blocks have quantum mechanical origin, or more specifically the presence of $\hbar$. Yet all the $\hbar$'s just get subtly cancelled and at the end a purely classical Newton's law has emerged. So one naturally wonders: where is $\hbar$ hiding in entropic gravity? 

There have been previous works aiming at finding the entropic corrections to Newton's law but however unsatisfactory. We will discuss with them in later sections.

We here argue that as there have already existed a minimal length scale in the entropic derivation, the generalized uncertainty principle (GUP) should be taken into consideration under this minimal length. Indeed based on GUP it has been demonstrated that the black hole Bekenstein entropy area law must be modified not only in the strong but also in the weak gravity regime \cite{GUPPisin}. In the weak gravity limit, such a GUP modified Bekenstein entropy exhibits a logarithmic correction. Such a log-correction is consistent with similar conclusions drawn from string theory, AdS/CFT correspondence, and loop quantum gravity considerations \cite{String,AdSCFT,LQG1,LQG2}. When applying it to the entropic derivations, we demonstrate that the resulting entropic gravity does include sub-leading order correction terms that depend on $\hbar$.

The organization of this paper is as follows. To address the question we posted, we first set up the key ingredients of entropic gravity framework toward the derivation of weak-field-limit gravity force law in section II, where the concept of entanglement entropy is introduced and the entropic derivation of Newton's gravitation force law reformulated in a self-consistent approach. Also in this section the entropy content in the system is clarified,  the entropy variation law re-derived, and a new reasoning for BH temperature based not on the equipartition theorem but on the blackbody radiation is introduced.
In section III we invoke the generalized uncertainty principle, which leads to a corrected form of black hole temperature and entropy. The modification of the information content so provided by GUP will result in the revision of the force law.
In section IV we briefly review and comment on some previous efforts aiming at finding the quantum corrections-- that is, the missing $\hbar$'s-- in the entropic gravity scenario. We then repeat the steps of Verlinde's, but with the entropy variation law and the temperature redefined by the GUP corrected entropy. We arrive at an exact force law of gravity at the end, and this exact force law recovers not only the classical Newton's law but also the sub-leading order quantum correction terms in the weak-field limit.
In section V, conclusions and comments are made about the implications of our findings. We suggest that the resulting deviation from the classical Newton's law may serve as a probe to the validity of the entropic gravity postulate.

\section{Entropic Derivation of Newton's Gravity}

In this section we will reformulate the derivation of classical Newton's law of gravity through thermodynamics and the leading behavior of holographic entanglement entropy. Some formulas will resemble that invoked by Verlinde in section 3.2 of Ref.\cite{Verlinde}, but the rationale behind his and ours are different. Our most crucial departure from Verlinde's previous work is that in our perspective, this entropic gravity approach is not an emergent description of gravity. This point will be elaborated more in the last section.

\subsection{Holographic entanglement entropy}
In the derivation of the entropic gravity force law, we believe that one should invoke the concept of entanglement entropy \cite{Information,EE,QE,EE&QFT,Susskind2} to find the information content of the system.
The entanglement entropy is a quantum mechanical quantity that measures the correlation between a subsystem A and its complementary subsystem B. When the world is divided into two subsystems, the total Hilbert space can be written as $\mathcal{H}_{\text{tot}}=\mathcal{H}_A\otimes \mathcal{H}_B$. If an observer can access the entire system, then the total entropy of the system is the quantum version of the classical Gibb's entropy, $S=-k_B\sum _i P_i \ln  \left(P_i\right)$, here $P_i$ the probability for a given state $i$, i.e., the von Newmann entropy for a statistical state in $\mathcal{H}_{\text{tot}}$ with density matrix $\rho_{tot}$: $S(\rho _{tot})= -k_B\text{Tr}(\rho  _{tot}\ln  \rho _{tot})$ \cite{Information}. For an observer who can only access the information of subsystem A, she will feel as if the state is described by a reduced density matrix $\rho _A=\text{Tr}_B\rho _{\text{tot}}$, where the trace is a partial trace over all eigenstates in $\mathcal{H}_B$ for the total density matrix. The entanglement entropy is thus defined as the von Neumann entropy for the reduced density matrix $\rho _A$: $S_A= -k_B\text{Tr}_A\left(\rho _A \text{ln$\rho $}_A\right)$. If the total state is entangled, that is, if it is not factorizable as $\left.\left.|\Psi _{tot} \rangle =\left|\Psi _A\right.\right\rangle \otimes |\Psi _B\right\rangle $, then the entanglement entropy is non-vanishing even if the total state is a pure state with zero entropy \cite{EE}. 

It can be shown by straight-forward calculations that the entanglement entropy of subsystem A is equal to that of subsystem B if the total state is pure \cite{EE}. Srednicki \cite{Srednicki} pointed out that with the property $S_A=S_B$, the entanglement entropy for a pure state, which we often referred to as the unique ground state of the total system, should only depend on the properties shared by the two regions. Therefore, it is expected that the leading behavior for pure ground state of a quantum field system scales as the boundary area rather than the volume of the subsystems. This area-scaling leading behavior of the entanglement entropy has been revealed in various physical systems such as the quantum critical phenomena \cite{F.critical}, explicit calculations of quantum field systems \cite{EE&BB}, and the AdS/CFT correspondence in string theory \cite{AHEE,DHEE}. 
This area-scaling property of entanglement entropy is referred to as the holographic entanglement entropy: for a quantum field theory in a space that is divided by a surface $\Sigma$ into two regions, the entanglement entropy for the ground state of the field is
\begin{align}
S_E=\frac{\text{Area}(\Sigma ) c^3k_B }{4\hbar  G}+ {\text{subleading terms}}.
   \label{HG3:entangle}
\end{align}%(3)
Here Area$(\Sigma)$ is the area of the surface, and $G$ is Newton's constant (see also \cite{HEEoverview} for a review).

The condition of the holographic entanglement entropy has been demonstrated for minimal surfaces with vanishing extrinsic curvature \cite{EE&BB,F.critical,AHEE,DHEE}. Some special cases of nonvanishing extrinsic curvature such as 2-sphere and 2-d cylinder also possess this property \cite{EESphere}
The entanglement entropy can be renormalized by fixing the cutoff length of the theory at Planck Length $L_p$. Because this entropy of entanglement is associated with the quantum ground state, some refer to it as the entropy of the fundamental degrees of freedom for the underlying quantum field theory across the boundary, others may call it the entanglement entropy on the boundary surface.
 
The original motivation for the entanglement entropy was to give a statistical explanation for Bekenstein entropy in black hole thermodynamics. The entanglement entropy in quantum gravity has been known as the quantum corrections to black hole entropy from matter fields \cite{HEEoverview,EE&BB,BHJacob,Susskind}. Some further pointed out that the black hole entropy is a pure entanglement entropy if the entire gravitational action is `induced' by the quantum fluctuations inside and outside the event horizon \cite{HEEoverview,BHJacob,Susskind}. Thus the black hole entropy is provided by this correlation between the degrees of freedoms on opposite sides of the horizon. An observer outside the event horizon without the access to what happens inside will experience a thermal state associated to this entanglement entropy.

We should note that the entanglement entropy is not exactly proportional to the area; only the leading order term follows the Bekenstein's law: $S= A \left/4L_p^2\right.$. The correction terms for the entanglement will be discuss later in section III, and the fact that simple entropy-area relation is only valid in the leading order will be emphasized to retrieve the missing $\hbar$ factor in weak field entropic gravity hypothesis. In this section we will only treat the entropy following Bekebstein's law without any extra terms, as is the case in Verlinde's scenario.

\subsection{Entropic gravity scenario with Bekenstein form of entropy}

\subsubsection{Entropic gravity system}

In the entropic interpretation of gravity, a spherical screen is invoked with radius $R$ that centers at a massive source $M$ and separates the universe into two regions, one inside the sphere and the other outside. A test particle with mass $m$ is placed just outside the spherical screen, see FIG. 1. The spirit of this entropic gravity system is that for the test particle outside the sphere, it will interact thermodynamically with the screen on which the information of the massive source is registered. If the variation in the entropy occurs as the test particle moves, the test particle will then confronted a restoring force according to the first law of thermodynamics: $Fdx=TdS$. To find the form of this restoring force caused by the system's tendency toward the maximization of entropy, one first has to know how the entropy varies in response to the displacement of the test particle. If the temperature can also be determined, then putting these together one can arrive at the entropic force law. We will show that if the entanglement entropy on the sphere is normalized by Planck Length and we consider only its leading order behavior following the Bekenstein law, the restoring force will have the same form of Newton's law of gravity in the end.

We note here that Verlinde called this spherical screen a `holographic screen', on which the information content obeys the holographic principle so that the information inside the screen is registered by the number of bits that is proportional to its surface area. The use of the holographic principle here, however, is ambiguous or even misleading because the holographic principle only suggests an inequality in the information content \cite{'t Hooft,Susskind,Bousso}. According to this principle only the black hole horizon would saturate the upper-bound of the inequality and recovers Bekenstein's area law. Under this light, it would be more appropriate to refer to this assignment as                                                                                                                                                                                                                                                                                                                                                                                                          `the holographic formula for entanglement entropy'.
%We believe what Verlinde really intended to convey is that the microscopic degrees of freedom can be represented holographically on the boundary. Thus the appropriate terminology should be `the holographic formula for entanglement entropy' rather than `the holographic principle' itself. We therefore clarify the meaning of `holographic screen' as the `surface holding the holographic entanglement entropy property defined by Eq.(\ref{HG3:entangle}) in the quantum gravity spacetime', and the number of degrees of freedom of the system is proportional to the area of this surface. The entropy on the screen is the entanglement entropy associated with the separation of the spacetime regions defined by this screen.

\subsubsection{Entropy variation law}

\begin{figure} %FIG.1
\center
\includegraphics[scale=0.5]{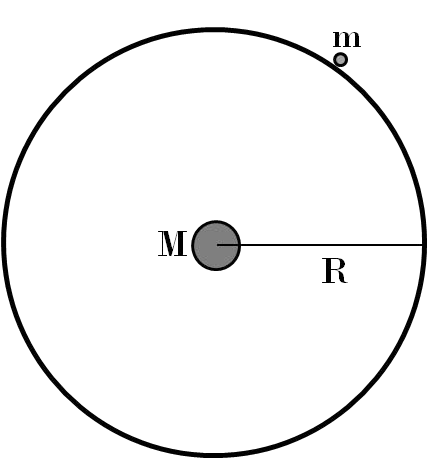}
\caption{Verlinde's system: a massive source $M$ is encoded by a spherical screen with radius $R$, and test particle $m$ is placed just outside the screen.} 
\end{figure}

With the system set up, we now proceed to see how the entanglement entropy changes as the test particle moves.

\begin{figure}  %FIG.3
\center
\includegraphics[scale=0.5]{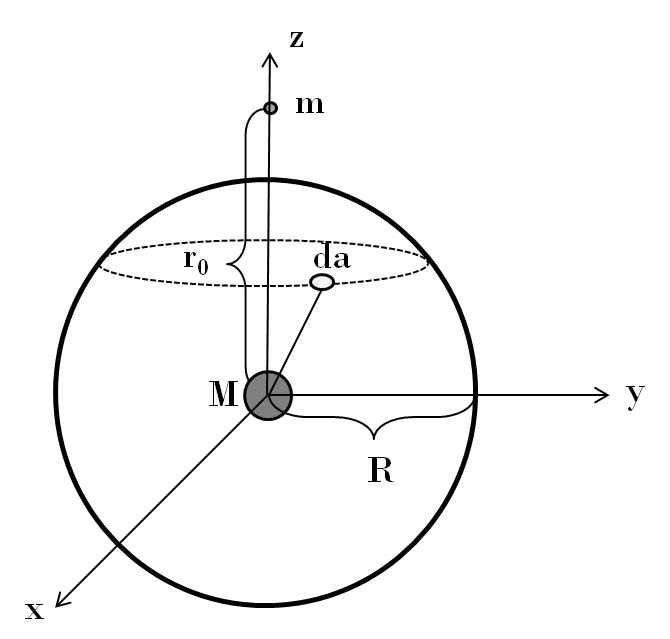}
\caption{Our system: a massive source $M$ located at the origin is encoded by a spherical screen with radius $R$. A test particle $m$ is placed outside the screen at a distance $r_0$ from the origin.}
\end{figure}

Consider a gravitational source of mass $M$ located at $r=0$. The spacetime metric in the weak field approximation is
\begin{align}
{ds}^2=-&\left(1-\frac{2GM}{r c^2}\right)c^2 {dt}^2+\left(1+\frac{2 G M}{r c^2}\right) {dr}^2
+r^2{d \Omega }^2
\notag\\
=-&\left(1-\frac{2 G M}{\rho  c^2}\right)c^2 {dt}^2+\left(1+ \frac{2 G M}{\rho  c^2} \right)\left({d \rho }^2+ \rho ^2 {d \Omega }^2 \right),
   \label{A1:metric}
\end{align}%(12)
where $\rho =r\left(1-G M/r c^2\right)$ and $d{\Omega}^2=d{\theta}^2+\sin^2\theta d{\phi}^2$.

In the system of interest, there is a massive source $M$ located at $(x,y,z)=(0,0,0)$ and a test particle $m$ located at $(0,0,r_0)$. A 2-sphere with radius $r=R$ surrounding $M$ is a surface that possesses the holographic property of entanglement entropy, which partitions the universe into two complementary regions to which $M$ and $m$ separately belong (see FIG.3).
The area of the surface no longer equals to $4 \pi R^2$ because of the slight warpage of the metric induced by the presence of the test particle. The metric in this system becomes

\begin{align}
&{ds}^2= -\left(1-\frac{2 G M}{\rho c^2}- \frac{2 G m/c^2}{ \sqrt{ {\rho _0}^2+\rho ^2-2 \rho _0 \rho \cos \theta }}\right)c^2 {dt}^2
\notag\\
&+ \left(1+\frac{2 G M}{\rho  c^2}+\frac{2G m /c^2}{  \sqrt{ {\rho _0}^2+ \rho ^2- 2 \rho _0 \rho \cos \theta }}\right)\left( {d \rho}^2+ \rho ^2 {d \Omega }^2 \right),
  \label{A2:metric}
\end{align}%(13)
where $\rho =R\left(1-G M/R c^2\right)$ and $\rho _0=r_0\left(1-G M/r_0c^2\right)$.

The surface area of the sphere is therefore 
\begin{align}
A=&\int \rho ^2 \sin \theta  d \theta   d \phi \left(1-\frac{2 G M}{c^2\rho}-\frac{2G m/ c^2}{ \sqrt{\rho _0{}^2+\rho ^2-2 \rho _0 \rho  \cos \theta }}\right).
   \label{A3:area}
\end{align}%(14)

Keeping the leading order in $G m R/c^2$ and $G M R/c^2$, we find
\begin{align}
&r_0>R: A=4\pi  R^2-\frac{8\pi  G M R }{ c^2}+\frac{8\pi  G m R^2}{ c^2r_0}\ ,
\notag\\
&r_0<R: A=4\pi  R^2-\frac{8\pi  G M R }{ c^2}-\frac{8\pi  G m R}{ c^2}\ .
   \label{A5:A result}
\end{align}%(16)

We see that while to the leading order the surface area $A$ is equal to $4 \pi R^2$, its correction induced by the presence of the test particle at $r_0$ is contributed by the $8\pi  G m R^2/ c^2r_0$ term. (Here we assume that the test particle is outside the sphere.) Now we like to see how an infinitesimal displacement of the test particle $m$ would further affect the surface area of the sphere. 
When the test particle makes a small displacement $\Delta r_0$ away from the sphere, the area will change by an amount
\begin{align}
\frac{\partial A}{\partial r_0}\Delta  r_{0 }= -\frac{8\pi  G m R^2}{ c^2r_0{}^2}\Delta r_0 \ .
   \label{A6:delta A}
\end{align}%(17)
Therefore if the entropy on the holographic screen follows the Bekenstein's law, then the entropy variation induced by the displacement of the test particle should be
\begin{align}
\Delta  S=k_B\frac{\Delta  A}{4l_p{}^2}= -\frac{2\pi k_B R^2 }{ r_0{}^2 }\frac{m c}{\hbar } \Delta  r_{0 }\ .
   \label{A7:delta S}
\end{align}%(18)
When the test particle is just outside the sphere, that is,
$R\approx r_0$, but with $R- r_0\gg G m/c^2$ to satisfy the weak field condition, the entropy variation on the sphere becomes
\begin{align}
\Delta  S=-2\pi k_B \frac{m c}{\hbar } \Delta  r_{0 } .
   \label{A8:delta S1}
\end{align}%(19)

This entropy variation law coincides with that conjectured by Verlinde. Verlinde's argument relies on the quantum uncertainty of a particle's position, that is, the position of a particle is indistinguishable within one Compton wavelength from the horizon. Since the precise location of the particle is unresolved within one Compton wavelength, how, therefore, would the horizon be able to react to the infinitesimal displacement within this uncertainty?

A more critical issue of Verlinde's argument toward the entropy variation law has to do with the possible inconsistency in his approach. There are two equations corresponding to the nature of entropy. One is the entropy variation law and the other is the Bekenstein law: $S= A \left/4 \right. L_p^2$, which was implicitly used through the holographic formulation of entanglement. 

Prior to Verlinde's conjecture, there existed a metric calculation of entropy variation law given by Fursaev \cite{FMinimal,FPlateau}. Fursaev used two infinite surfaces as the screens to divide the spacetime. However his derivation is only valid for the special case of infinite surfaces. We believe that a sphere is a physically more suitable geometry, since one can introduce a uniform temperature more naturally on a sphere than a infinite surface.
Our work therefore follows Fursaev's approach but apply it to the variation of the surface area of a sphere. Through that we manage to reproduce the entropy variation law suggested by Verlinde without the need to invoke his ambiguous Compton wavelength argument.

\subsubsection{Temperature}

Once the entropy variation associated with the displacement of the test particle is established, the only remaining task is to define the temperature as the final step towards the entropic gravity force law. Here we suggest a heuristic derivation of Hawking temperature for Schwarzschild black hole in terms of its mass and entanglement entropy. In terms of black hole thermodynamics, the Hawking temperature can be viewed as the blackbody radiation temperature associated with its evaporation. In this regard, the averaged energy is $~2.7 k_B T$ per photon based on statistical mechanics. The degrees of freedom for a black hole is $N=S_B/k_B$ (see, for example, Ref.\cite{Bousso}, such argument based on holographic principle). We suppose that these degrees of freedom, N, are associated with the number of the blackbody photons and that the total energy of the blackbody radiation in turn takes up the entire rest mass energy $Mc^2$ of the BH. Thus the temperature can be written as
\begin{align}
T=\frac{M c^2}{a S_B}\ ,
   \label{T1}
\end{align}%(20)
where $a$ is a constant of order one. To reproduce the correct form of Hawking temperature, we fix the coefficient and arrive at a form for the temperature following Bekenstein's law of BH entropy:
\begin{align}
T=\frac{M c^2}{2 S_B}=\frac{2G\hbar}{c k_B}\frac{M}{A}\ .
   \label{T1:a}
\end{align}%(20)

We now arrive at the same form of temperature as proposed by Verlinde. However, his derivation invokes the equipartition theorem of energy, which is a classical concept and is therefore unjustified in the entropic gravity framework, where the entropy has already involved $\hbar$, an indication of the quantum nature of the formulation.

With $T$ and the relation between $\Delta S$ and $\Delta x$ fixed, we are now ready to  the form of entropy.
When the test particle makes a small displacement $\Delta x$ relative to the screen, the entropy on the screen will change by an amount $\Delta S$ according to Eq.(\ref{A8:delta S1}). The test particle will therefore experience a restoring force originated from the system's tendency to increase its entropy. Unlike the restoring force of a stretched polymer which has two possible directions due to a finite nonvanishing equilibrium position, the entropic gravity force has only one direction, which corresponds to bringing two massive objects closer to each other. This ``entropic force law'' should thus follow the first law of thermodynamics:
\begin{align}
F  \Delta x=T \Delta S\ .
   \label{T5:1stloftherm}
\end{align}%(23)
Following Eqs.(\ref{A8:delta S1})--(\ref{T5:1stloftherm}) and equating the area of the spherical screen to $4 \pi  R^2 $ in the leading order approximation, we finally obtain the entropic force law that is identical to Newton's force law of gravity,
\begin{align}
F=-\frac{G M m}{R^2}\ .
   \label{T6:N's Grav}
\end{align}%(24)
The minus sign in this force law indicates that the entropic force is oriented opposite to the direction of the displacement, just as in Newton's view of the gravitational force that is attractive between two massive sources. 

While Newton's force law of gravitation seems to emerge elegantly through this entropic reasoning, we should emphasize again that both the entropy variation formula and the temperature formula involve an $\hbar$, which manifests their quantum origin. Both these two $\hbar$'s are originated from the information content of the holographic screen, where one comes out of the direct calculation of the entropy formula and the other emerges from the distribution of the degrees of freedom on the surface. The complete cancellation between these two $\hbar$'s was due to the coincidence that both the degrees of freedom, $N$, and the Bekenstein law are straight-forwardly proportional to the surface area of the holographic screen, which was fortuitous. We will argue in the next section that the entropy of entanglement is not exactly proportional to the area. As demonstrated in Ref.\cite{GUPPisin}, the generalized uncertainty principle (GUP) implies a corrected formula for entanglement entropy not only in the strong gravity but also in the weak gravity regime. 

In the derivation of the entropic gravity, the actual form of the entropy is a key ingredient. Extra care must therefore be taken in the determination of the BH entropy. With this in mind we emphasize that the holographic formulation of entanglement entropy is based on a cutoff length of the same order of the Planck length. This introduction of the cutoff length implies the existence of a minimal length scale that is essential in the entropic interpretation of gravitational force. The standard Heisenberg uncertainty principle, which is deduced under the Minkowski spacetime, must be modified, or generalized, when the spacetime cannot be reduced indefinitely but is subject to some minimal length scale \cite{Hossenfelder}. Originally suggested in 1960s \cite{Mead} based purely on the considerations of GR, GUP acquires additional theoretical support from string theory's perspective \cite{Veneziano,Gross,Amati,Konishi,Witten} since 1980s.

\section{Generalized Uncertainty Principle}
 One important implication of GUP is that the standard forms of Bekenstein entropy and Hawking temperature no longer hold as the size of a black hole approaches the Planck length \cite{GUPPisin}. A direct consequence of this GUP modified BH entropy is that the BH evaporation process will come to a stop when its Schwarzschild radius approaches the Planck length. As a result the Hawking evaporation should leave behind a BH remnant at Planck mass and size.

Based on GUP, it was found that the modified BH temperature is of the form \cite{GUPPisin}
\begin{align}
T_{GUP}=\frac{M c^2}{4\pi k_B}\left[1-\sqrt{1-\frac{M_p{}^2}{M^2}}\right]
   \label{G1:TGUP}
\end{align}%(25)
for a Schwarzschild black hole of mass $M$. In the large mass limit, i.e., $M_P / M \ll 1$, the BH temperature is
\begin{align}
T_{GUP}=\frac{c^2 M_P{}^2}{8 \pi  k_B M} \left[1+\frac{M_P{}^2}{4M^2}+\frac{M_P{}^4}{8M^4}+\text{...}\right]\ ,
   \label{G2:TGUPE}
\end{align}%(26)
which agrees with the standard Hawking temperature at the leading order. 

Since the black hole temperature has been modified, the entropy obtained by $S=\int dMc^2/T $ must also be correspondingly modified to a form different from the simple Bekenstein entropy expression: 
\begin{align}
S_{GUP}=2 \pi k_B \Bigg\{\frac{M^2}{M_p{}^2}\left(1-\frac{M_p{}^2}{M^2}
+\sqrt{1-\frac{M_p{}^2}{M^2}}\right)\notag\\
-\log \left[\frac{M}{M_p}\left(1+\sqrt{1-\frac{M_p{}^2}{M^2}}\right)\right]\Bigg\}\ .
   \label{G3:SGUP}
\end{align}%(27)
The integration constant of the integral is fixed by setting the entropy to zero at the final remnant state.
Thus in the large mass limit we have
\begin{align}
S_{GUP}=&4 \pi  k_B\frac{M^2}{M_p{}^2}-\pi  k_B \log\left(\frac{M^2}{M_p{}^2}\right)+ \text{const.} +\text{...},
\notag\\=&k_B\frac{A}{4L_p{}^2}-\pi  k_B \log\left(\frac{A}{L_p{}^2}\right)+\text{const}.+\text{...}\ ,
   \label{G4:SGUPE}
\end{align}%(28)
which recovers Bekenstein entropy as $M_P/M$ goes to zero.

The correction to the semiclassical area law of black hole entropy has been extensively studied. For example a generic logarithmic term as the leading correction to black hole entropy has been found universal up to a coefficient of order unity based on string theory and loop quantum gravity considerations, see for example \cite{String,AdSCFT,LQG1,LQG2}. Such logarithmic correction also appears in the entanglement entropy on minimal surfaces and some special nonminimal cases such as the 2-sphere in flat space \cite{EESphere}.
Here we treat GUP as a basic assumption to provide the correct form of holographic entanglement entropy because of its fundamentalness when dealing with issues related to quantum gravity. It is interesting to note that the GUP-based correction to the entanglement entropy has its IR limit that is in agreement with the well-supported logarithmic sub-leading corrections deduced from string theory or loop quantum gravity. However a fundamental difference between the GUP and other approaches is that the GUP correction to the entanglement entropy as shown in Eq.(\ref{G3:SGUP}) is an exact form, valid for both the UV and the IR limits. Therefore this GUP corrected form of entropy is also valid in the UV limit, which will be useful in our future work to extend our result to the strong gravity regime.

As the BH entropy is precisely the entanglement entropy on the BH horizon, we assert that under GUP the area-dependence of entanglement entropy is now expressed in the correct form as
\begin{align}
S_{GUP}=&\frac{A k_B}{8L_p{}^2}\left[1-\frac{16 \pi  L_p{}^2}{A}+\sqrt{1-\frac{16 \pi  L_p{}^2}{A}}\right]
\notag\\-&2 \pi  k_B \log\left[\frac{A}{4 \sqrt{\pi }L_p}\left(1+\sqrt{1-\frac{16 \pi  L_p{}^2}{A}}\right)\right]\ ,
   \label{G5:SGUPA}
\end{align}%(29)
which reduced to Eq.(\ref{G4:SGUPE}) in the large BH mass limit.

\section{Quantum Effects in Entropic Gravity}

\subsection{Quantum corrections to entropic gravity: other approaches}

There have been previous efforts aiming at finding the entropic corrections to Newton's law but however unsatisfactory. 
Santos et al. \cite{Santos} used the uncertainty principle to postulate a corrected entropy variation law and obtained the uncertainty in Newton's law of gravity. Ghosh \cite{Ghosh} further extended the previous work by using the idea of GUP. However, their approaches led to a force law that depends on the uncertainty in position, which is bothersome.  

Modesto et al. \cite{Modesto} introduced a log correction and a volume-scale correction term to the area-entropy relation, and arrived at a modified gravitation force law. We here point out that when the entropy-area law is changed, the number of bits is no longer simply inverse proportional to the Planck area. As a consequence, the temperature defined by equipartition rule will also be modified. Modesto et al. considered only the corrections to entropic variation law without noticing that the form of temperature should also be corrected. Setare et al. \cite{Setare} revisited Modesto's idea and modified Newton's gravitational force law via GUP and self-gravitational corrections. Setare et al. modified both entropy and temperature but failed to introduce, in our opinion, the right form of GUP-modified entropy, which should follow the well-acknowledged logarithmic form of correction. Another thing we should note here is that both Modesto et al. and Setare et al. suggested only the sub-leading corrections of the force law rather than an exact form.
Nicolini \cite{Nicolini} obtained an exact form of the corrected entropic force law via noncommutative gravity and ungraviton corrections, but he also did not take into consideration the effect of modified information content on the definition of the temperature.

Here we argue that one can trace the quantum effects in the entropic gravity scenario by invoking an exact form of GUP corrected entropy formula. This GUP corrected entropy formula has a universally-accepted logarithmic leading-order correction, and its deviation from the Bekenstein law will affect the entropic variation equation as well as the temperature and therefore will lead us to a quantum corrected gravitational force law.

\subsection{Entropic gravity under GUP}

In Verinde's entropic gravity scenario, the purely classical Newton's force law of gravitation is derived based on a quantum-mechanical and thermodynamical setup. To keep track of the underlying quantum dynamics, we now invoke generalized uncertainty principle to uncover the missing quantum contribution in entropic gravity.

%Under the influence of GUP, how entanglement entropy depends on boundary surface area is corrected to a relation different from the exact proportionality suggested in Bekenstein law. Therefore, the information content on the holographic screen changes, and thermodynamic around the screen is now different from the outcome of Verlinde's consideration in only the leading behavior for large distance. 

Again we consider a spherical holographic screen, whose information content is defined by the GUP corrected entanglement entropy, encoding a massive source $M$ at the center and a test particle $m$ placed just outside this spherical surface of radius $R$. The restoring force acting on the test particle $m$ induced by the displacement from its (equilibrium) location will be derived based on the first law of thermodynamics. 

First of all, the entropy variation law is directly affected by the GUP corrected form. Under GUP, the entropy varies with the surface area as
\begin{align}
\Delta S=\frac{\partial S_{GUP}}{\partial A} \Delta A\ ,
   \label{C1:delataSGUP}
\end{align}%(30)
with $\Delta A=-8 \pi  G m / c^2$ as calculated before.

Next we determine the temperature on the screen. Under the GUP framework, the number of bits on the screen will become
\begin{align}
N=\frac{S_{GUP}}{k_B}\ .
   \label{C2:NGUP}
\end{align}%(31)
We again apply the same argument based on the mean energy of blackbody photons and connect it with the counting of degrees of freedom on the screen to arrive at the following form of temperature:
\begin{align}
T=\frac{2M c^2}{N k_B}=\frac{M c^2}{2 S_{GUP}}\ .
   \label{C3:Tcorrected}
\end{align}%(32)
Finally, using the first law of thermodynamics we arrive at the modified gravity force law:
\begin{align}
F_{GUP}=F_N \frac{2((1+\eta )-2\alpha  (2+\eta ))}{ \eta  (1+\eta ) \left(-4\alpha +(1+\eta )+4 \alpha  \log \left[\frac{2\sqrt{\alpha }}{1+\eta }\right]\right)}.
   \label{C4:FGUP}
   \end{align}%(33)
Here $F_N=GmM/R^2$ is Newton's gravitational force law, and we have introduced symbols $\eta =\sqrt{1-4G\hbar/c^3R^2}$ and 
$\alpha =G\hbar/c^3R^2$ to simplify the expression.

In the large distance limit where $R\gg  L_p=\sqrt{G\hbar/c^3}$ and therefore $\alpha =G \hbar/c^3R^2 \ll 1$, we can expand the force to the third order of $\alpha $ as
\begin{align}
F_{GUP}=&F_N\{1+\alpha[2- \log\alpha]+\alpha^2[4-5\log\alpha +(\log\alpha)^2]\notag\\ 
&+\alpha^3[7-18\log\alpha +8(\log\alpha)^2-(\log\alpha)^3]+\text{...}\}\ .
\label{C5:FGUP3}
\end{align}%(34)
It is clear that this GUP-based force law recovers the classical Newton's gravitational force law in the infinite distance limit, while some subleading quantum corrections is present as long as $\alpha$ is finite. On the other hand these correction terms go to zero in the classical limit as $\hbar$ vanishes. These $\alpha$-dependent terms, we conclude, are where $\hbar$ is hiding in entropic gravity.
%In the opposite limit when the value of holographic screen radius $R$, also the distance in Newton's law, reaches the Planck scale, the force goes to infinity. However, in this extreme short distance limit the force law may have to be modified again base on extremely small black hole physics, and we will discuss about it in the next section.
   
\section{Conclusion and discussions}

In this paper we raised the question about where $\hbar$ is hiding in entropic gravity. Through the reanalysis of the fundamental building blocks of entropic gravity, in particular the holographic formulation of the entanglement entropy, we argued that the perfect cancellation of $\hbar$'s among all the quantum mechanically motivated inputs is broken if the more exact form of the BH entanglement entropy based on GUP is to replace the Bekenstein area law. Based on this we found, in the weak gravity limit, the hided $\hbar$'s in the form of logarithmic corrections to the classical Newton's law, in Eq.(\ref{C5:FGUP3}).

In our attempt of seeking the missing $\hbar$'s, we reformulated the existing derivations of entropic gravity. In Verlinde's entropic gravity derivation, two ingredients involving entropy formula have been invoked without the guarantee of their mutual compatibility. 
We applied Fursaev's procedure to reproduce the leading order entropy variation in Verlinde's setup of spherical holographic screen.

While our approach manages to avoid the compatibility issue, there is a price to pay. In our alternative approach we have introduced the concept of spacetime metric and its deformation due to the presence of a massive object, which implicitly assumed the knowledge of general relativity, the standard theory of classical gravity. Yet the very attempt of entropic gravity is to deduce it from quantum mechanics and statistical physics alone without any prior knowledge of gravity. We are therefore at risk of a circular logic in our approach if gravity is to be interpreted as an emergent phenomenon. In this regard a more cogent and consistent argument without involving any gravity-related concept is needed towards an alternative entropy variation law, in order to assert the validity of the entropic framework of gravity as an emergent phenomena.  
By the similar token, the existing derivations of entropic gravity also faces the similar issue since Newton's constant has been invoked as a fundamental constant from the outset instead of being a secondary, derived parameter of the theory as it should if gravity is to be an emergent phenomenon. 

Under this light one can instead view our derivation of the entropic gravity not as an emergent phenomenon but as a means to deduce the `quantum gravity force law' via the quantization of the information content on the surfaces in units of Planck area provided by GUP as well as the spacetime warpage effect in the presence of a massive particle provided by general relativity.

Although there are still rooms to improve in this line of approach to gravity, we have provided an exact form of quantum corrected entropic gravity force law based on the assumption of GUP as a fundamental input. Such quantum corrections, though minute, may serve as a probe to examine the concreteness of the entropic gravity interpretation in the the experimentally measurable scale of large distance and weak gravity limit. 

\section*{Acknowledgement}
We thank Debaprasad Maity, Taotao Qiu, Keisuke Izumi, Yen-Chin Ong , Chien-I Chiang, Nian-An Tung and Jo-Yu Kao for helpful and inspiring discussions. This research is supported by the Taiwan National Science Council (NSC) under Project No. NSC98-2811-M-002-501, No. NSC98-2119-M-002-001, and the US Department of Energy under Contract No. DE-AC03-76SF00515. We would also like to thank the NTU Leung Center for Cosmology and Particle Astrophysics for its support.

\end{document}